\documentclass[12pt]{article}
\usepackage{colortbl}
\usepackage{verbatim}
\usepackage{amssymb,amsfonts,amsmath,amsthm}
\usepackage{geometry}
\usepackage{alltt}
\usepackage{multirow}
\usepackage{graphicx}
\usepackage{subfigure}
\usepackage{setspace}
\usepackage{xcolor}
\usepackage{soul}
\usepackage{breqn}

\usepackage{booktabs} 
\usepackage{rotating}
\usepackage{bigstrut}

\onehalfspace

\usepackage{lscape}  
\usepackage{array} 
\usepackage[english]{babel}
\usepackage[sort, authoryear]{natbib} 
\usepackage{url}
\makeatletter
\renewcommand\@biblabel[1]{#1.} 
\makeatother

\usepackage{bigstrut}
\usepackage[colorlinks,citecolor=blue,urlcolor=blue]{hyperref}
\usepackage{subfigure}
\usepackage{multirow}

\begin{document}
\title{Estimation of logistic regression parameters for complex survey data: a real data based simulation study}

\author{Amaia Iparragirre \footnote{Corresponding author: {{E-mail: amaia.iparragirre@ehu.eus}}, Address: Departamento de Matem\'aticas. Universidad del Pa\'is Vasco UPV/EHU.} $^{1}$\\
        Irantzu Barrio$^{1,3}$\\
        Jorge Aramendi$^{2}$ \\
        Inmaculada Arostegui$^{1,3}$\\		      
        \small{$^{1}$ Departamento de Matem\'aticas.} 
        \small{Universidad del Pa\'is Vasco UPV/EHU}\\
        \small{$^{2}$ Eustat - Euskal Estatistika Erakundea - Instituto Vasco de Estad\'istica}\\ 
        \small{$^{3}$ BCAM - Basque Center for Applied Mathematics}\\
}

\date{}
\maketitle

\begin{abstract}
 
In complex survey data, each sampled observation has assigned a sampling weight, indicating the number of units that it represents in the population. Whether sampling weights should or not be considered in the estimation process of model parameters is a question that still continues to generate much discussion among researchers in different fields. We aim to contribute to this debate by means of a real data based simulation study in the framework of logistic regression models. In order to study their performance, three methods have been considered for estimating the coefficients of the logistic regression model: a) the unweighted model, b) the weighted model, and c) the unweighted mixed model. The results suggest the use of the weighted logistic regression model, showing the importance of using sampling weights in the estimation of the model parameters.

\end{abstract}


\section{Introduction}\label{Introduction}

Complex survey data are gaining popularity among researchers and analysts from different fields. In particular, complex survey data are usually used, among other purposes, to fit prediction models. These data are commonly obtained by sampling the finite population that is of interest for the survey by some complex sampling design. One of the characteristics of this type of data are sampling weights, which indicate the number of units that each sampled observation represents in the finite population. When working with complex survey data, before implementing the traditional statistical techniques, most of which have been designed to be implemented on simple random samples, it should be analyzed whether these techniques are valid to this kind of data \citep{Skinner1989}. 
 
In particular, whether or not to use the sampling weights when fitting prediction models is a question that has been widely discussed in the literature by a number of researchers \citep{Brewer1973,Smith1981}. Different perspectives can be adopted when fitting prediction models to survey data, which are usually denoted as model-and design-based approaches \citep{Binder2009,Chambers2003}. On the one hand, the researchers that adopt the design-based perspective warn that if the complex sampling design, and in particular, the sampling weights are not considered in the estimation process of model parameters, the variances tend to be underestimated and biased estimates may be obtained \citep{Binder2009,Heeringa2017}. Therefore, they claim that the sampling weights should be considered in the estimation process of model parameters.

On the other hand, from a model-based point of view, if the model is well specified the coefficient estimates must be unbiased even though the sampling weights are not considered directly in the estimation process and considering them may increase the standard deviations of the estimates, particularly for small sample sizes \citep{Scott1986,Reiter2005,Chambers2003,Korn1995}. In this context, \cite{Rubin1976,Scott1977,Sugden1984} established conditions under which the sampling design may be ignored for inference purposes. As explained by \cite{Skinner1989} a condition for a design to be ignorable is to be noninformative. A sampling design is denoted as informative if the response variable is related to the sampling weights, even after considering the covariates that are going to be part of the model \citep{Pfeffermann2009}. Different methods have been proposed from the model-based perspective in order to ensure that the design is ignorable and the models are well specified \citep{Pfeffermann2009}. Researchers that adopt this perspective propose, among other techniques, to incorporate into the model as covariates all the design variables that have been considered in the sampling process and the interactions between them (see, e.g., \cite{DeMets1977, Nathan1980, Gelman2007}). 

Although it was already pointed out by \cite{Chambers2003}, the discussion between the two perspectives is still alive. Some more recent works, such as \cite{Reiter2005}, \cite{Masood2016} and \cite{Lumley2017a}, show that this debate still generates doubts among researchers and makes it difficult to decide whether or not to use sampling weights in their analyses. Most researchers agree that it is not advisable to ignore sampling weights if the sample is informative or the model is not well specified, but at the same time, they encourage to ignore the sampling weights when they are not strictly necessary. The difficulty usually lies in identifying whether or not sampling weights are necessary to estimate model parameters based on our particular survey data, or put it another way, whether or not the design is informative. As explained by \cite{Pfeffermann2009}, informativeness depends not only on the sampling design, but also on the model that is going to be fitted, the response variable of that model and the covariates that will be included. Therefore, commonly it is not easy to know whether the sampling design of the survey data to be analyzed is informative or not to fit a particular model. In addition, it is not always possible to include all the design variables and the interactions between them in the model due to several reasons, such as the lack of information, the large number of design variables and the fact that when including design variables as covariates into the model it may lose scientific interpretability \citep{Pfeffermann2009}. Consequently, nowadays, it is not easy to decide in practice whether sampling weights should or not be considered for estimating model parameters yet. For this reason, we believe that further studies are needed in this area and, in particular, we consider that it is necessary to provide insight considering real data based simulation studies, as a complement to the theoretical results and case studies that have been most discussed so far. 

Throughout this work we focus on the estimation of model parameters and, in particular, on the logistic regression framework for dichotomous response variables. Although in general there are more studies concerning about the linear regression model (see, e.g., \cite{DeMets1977,Nathan1980,Holt1980,Hausman1981}), a number of works have also been carried out in order to address this problem arising from complex samplings in the field of logistic regression models. In particular, \cite{Scott1986, Scott2002} work with simulated data inspired on a case-control study. It should be noted that case-control studies consist on stratifying the data based on the dichotomous response variable, and therefore, are always based on informative sampling designs. But, what if we do not know whether our sampling design is informative or not to fit a certain model? As mentioned above, in practice, this is the situation that usually occurs when working with real complex survey data.  \cite{Chambless1985, Lumley2017a} and \cite{Reiter2005} raise this issue in their analysis with real survey data and they compare several estimation methods adopting both, model- and design-based perspectives and they finally select the most appropriate model for their analysis. However, how can we know in practice whether these differences in estimates are large or not, and if so, which of the estimates is the most appropriate? In this work we aim to go a step further and contribute to the work that has been done in the above-mentioned papers by analyzing the differences among different methods by means of a real data based simulation study, in order to work under a real-life scenario that allows us to compare the coefficient estimates to the theoretical ones. Hence, data were generated based on real surveys and, a priori, whether these data are informative or not to fit different models it is not known for us in advance. Our goal is to analyze by means of a simulation study a situation that frequently occurs in practice and to analyze and evaluate the consequences or the effect of making the decision to consider or not the sampling weights to estimate the coefficients of the logistic regression model in each situation. In this study we compare the performance of several estimation methods that are commonly applied for estimating the coefficients of the logistic regression model (see, e.g.,  \cite{Lumley2017a}). In particular, we compare the coefficient estimates obtained by: a) the unweighted logistic regression model, b) the weighted logistic regression model, and c) the unweighted logistic regression mixed model with random intercept. Different scenarios were defined based on a) two real survey data; and b) number of covariates/parameters in the model. The real surveys were designed and collected by the Official Statistic Basque Office (Eustat) based on one-step stratification with simple random sampling in each stratum. 

The rest of the document is organized as follows. In Section \ref{datasets} we describe the two original real surveys that motivated this work: ESIE and PRA surveys. In Section \ref{methods} the methods that were applied for estimating the model parameters are described. Information about the simulation procedure, scenarios that were drawn and the results we obtained can be found in Section \ref{simulationstudy}. In Section \ref{application}, we apply the described methods to real survey data for illustration purposes. Finally, the paper concludes with a discussion in Section \ref{discussion}. 

\section{Motivating data sets}\label{datasets}

In this section, we describe the two complex surveys that motivated this work. Those surveys were designed, conducted and collected by the Official Statistics Basque Office (Eustat). 

On the one hand, the Information Society Survey\footnote{\url{https://en.eustat.eus/estadisticas/tema_150/opt_1/tipo_7/temas.html}} in companies, which is usually denoted as ESIE survey for its Spanish acronym, was carried out among the companies in the Basque Country (BC) in order to collect information about the use of technology. In particular, the response variable that we concern about in this study is a dichotomous response variable that indicates whether a company has its own web-page (1) or not (0), which we aim to model by means of covariates such as ownership, activity or number of employees of the establishment. On the other hand, the Population in Relation to Activity (PRA) Survey\footnote{\url{https://en.eustat.eus/estadisticas/tema_37/opt_0/temas.html}} was conducted among the inhabitants of the BC aged 16 and over, with the aim of estimating the percentage of the labor force of the BC. Specifically, the response variable that we consider in this study indicates whether each individual is active (1) or not (0). Among the most important covariates we found age, educational level, nationality, and sex. 

In both surveys, the two finite populations were sampled based on one-step stratification with simple random sampling in each stratum, i.e., the populations were split into different strata, and a certain number of units (that were previously determined) were sampled randomly from each stratum. Nevertheless, the strata were defined in very different ways in both surveys. In the ESIE survey, strata were defined based on the combination of three categorical variables which are 1) province where the company is located (that takes 3 categories), 2) activity of the company (in 65 categories) and 3) number of employees (3 categories). Therefore, a large amount of small strata, a total of 585 
were defined. However, it should be noted that in some of these strata there are no units in the population, so in fact we have 515 strata in total. In contrast, in the PRA survey, strata are the 23 regions of the BC. This causes the response variable to be distributed differently in each stratum in ESIE, while in PRA, there are no differences of the distribution of the response variable among the strata. In both, ESIE and PRA surveys, once the sample was obtained from the finite population following the described sampling process, a sampling weight was assigned to each sampled unit.
 
In the ESIE survey, from the finite population of $195\,222$ companies, $7\,725$ were sampled (these data was collected in 2010). In particular, strata sizes in the finite population range from 1 to $14\,535$, where the median is 38 and the interquartile range $7-185.5$. The sampling probabilities for each stratum range from 0.0061 to 1, with a median of 0.2830 and an interquartile range of $0.0970-0.8417$. In contrast, in the PRA survey, from a total of $1\,851\,316$ individuals $10\,609$ were sampled (information related to the last quarter of 2016). Specifically, strata sizes range from $2\,768$ to $438\,595$, being the median $44\,335$ and $22\,024-72\,834$ the interquartile range. The sampling probabilities range from 0.0041 to 0.0488, with a median of 0.0063 (the interquartile range is $0.0055-0.0102$).

\section{Methods}\label{methods}

In this section, we describe the methods we have considered in order to estimate the logistic regression coefficients for complex survey data.

Let $Y$ indicate the dichotomous response variable, which takes the value 1 to indicate the event of interest (0 otherwise), and $\pmb X=(X_1,\ldots,X_p)^t$ the vector of $p$ covariates. Let $U=\left\{1,\ldots,N\right\}$ be a finite population for which $N$ realizations of the set of random variables $(Y,\pmb X)$ are associated, i.e., $\left\{(y_i,\pmb x_i)\right\}_{i=1}^N$. Let $S$ be a sample of $n$ observations drawn from the finite population $U$ by one-step stratification. Let $h=1,\ldots,H$ indicate the different strata. The sampling weights associated to each sampled unit $i\in S$ are denoted as $w_i$.

Let us define the true population logistic regression model as follows:
\begin{equation}
	logit(p_i)=\ln\left[p_i/(1-p_i)\right]=\pmb x_i^T\pmb\beta^{True}
\end{equation}
where $p_i=P(Y=1|\pmb x_i)$ denotes the probability of event for the unit $i$ given the values of covariates $\pmb x_i$ $(\forall i\in U)$ and the model coefficients $\pmb\beta^{True}=(\beta_0^{True},\beta_1^{True},\ldots,\beta_p^{True})^T$ are computed by maximizing the population likelihood:
\begin{equation}\label{population_likelihood}
	L_{pop}(\pmb\beta)=\prod_{i=1}^N p_i^{y_i}(1-p_i)^{1-y_i}.
\end{equation}

However, it should be noted that responses $y_i$ are usually known only for the sampled units, $i \in S$. For this reason, the model should be estimated based on the sample $S$. In this work, we compare the performance of several estimation methods that are commonly applied for estimating the coefficients of the logistic regression model for dichotomous response variable \citep{Lumley2017a}. The goal is to compare these estimates to $\pmb\beta^{True}$ in order to analyze the performance of each method.

In this context, a simple logistic regression model can be fitted to the complex survey sample $S$, which can be defined as follows:
\begin{equation}\label{logistic_regression_model}
\text{logit}(p_i)=\ln\left(\dfrac{p_i}{1-p_i}\right)=\pmb x_i^T\pmb\beta.
\end{equation}
Different methods can be applied to estimate the vector of regression coefficients $\pmb\beta=(\beta_0,\ldots,\beta_p)^T$ based on $S$:

\begin{itemize}
	
	\item[\textbf{M1.}] \textbf{Unweighted logistic regression model}
	
	This method consists in estimating the model coefficients by maximizing the likelihood function in equation \eqref{likelihood_M1} by means of some iterative algorithms such as the iteratively reweighted least squares (IRLS) algorithm \citep{McCullagh1989}: 
	\begin{equation}\label{likelihood_M1}
	L(\pmb\beta)=\prod_{i\in S}p_i^{y_i}\left(1-p_i\right)^{1-y_i}.
	\end{equation}
	Let us denote as $\hat{\pmb\beta}_{\text{M1}}$ the coefficients estimated by means of the maximum likelihood method, hereinafter.
	
	\item[\textbf{M2.}] \textbf{Weighted logistic regression model}
	
	This approach consists in estimating the coefficients that maximizes the pseudo-likelihood function in equation \eqref{likelihood_M2} \citep{Binder1981,Binder1983} which considers the sampling weights $w_i$:
	\begin{equation}\label{likelihood_M2} 
	PL(\pmb\beta)=\prod_{i\in S}p_i^{y_iw_i}\left(1-p_i\right)^{(1-y_i)w_i}.
	\end{equation}
	The pseudo-likelihood function is also maximized by means of iterative algorithms \citep{Heeringa2017,Wolter2007}. Let us denote as $\hat{\pmb\beta}_{\text{M2}}$ the coefficient estimates obtained based on this method.
	 
\end{itemize}

In addition to the above-mentioned methods, another option is to fit a mixed model considering the complex sampling design as second level units (see, e.g. \cite{Lumley2017a,Masood2016}). In this study, in particular, we consider a random intercept model in the same way as in \cite{Lumley2017a}. Let $i=1,\ldots,n_h$ indicate the sampled units belonging to stratum $h$ $(\forall h\in\{1,\ldots,H\})$, while $\pmb x_{hi}$ and $y_{hi}$ indicate the values of the vector of covariates and response variable for $i$ in stratum $h$, respectively. Then, we aim to fit the following model to our sample $S$:
\begin{equation}\label{logistic_random_intercpet}
	\text{logit}(p_{hi})=\ln\left(\dfrac{p_{hi}}{1-p_{hi}}\right)=\pmb x_{hi}^T\pmb\gamma + u_{h},\quad u_{h}\sim N(0,\sigma_{u}^2).
\end{equation}
where $p_{hi}=P(Y=1|\pmb x_{hi},u_{h})=\dfrac{e^{\pmb x_{hi}^T\pmb\gamma + u_{h}}}{1+e^{\pmb x_{hi}^T\pmb\gamma + u_{h}}}$. 

\begin{itemize}	

	\item[\textbf{M3.}] \textbf{Unweighted logistic regression model with random intercept}
	
	In this case, the likelihood function is defined as follows:
	\begin{equation}\label{likelihood_M3}
	L_{\text{mix}}(\pmb \gamma,\sigma_{u}^2)=\prod_{h=1}^H\int_{-\infty}^{+\infty} f(y_{hi}| \pmb x_{hi}, u_h)f(u_{h})du_{h},
	\end{equation}
	where $f(y_{hi}| \pmb x_{hi}, u_h)=\prod_{i=1}^{n_h}p_{hi}^{y_{hi}}(1-p_{hi})^{1-y_{hi}}$ and $f(u_h)=\dfrac{1}{\sigma_u\sqrt{2\pi}}e^{-{u_h^2}/{2\sigma_u^2}}$. The parameters $\pmb\gamma$ and $\sigma_{u}^2$ are commonly estimated by maximizing the likelihood function in \eqref{likelihood_M3} numerically, usually by means of Laplace approximation \citep{Lee2001}. Let us denote as $\hat{\pmb\gamma}$ and $\hat\sigma_{u}^2$ those estimates, respectively, hereinafter.
	
	However, the comparison of the coefficients obtained from conditional random effect models and the corresponding marginal models is not straightforward \citep{Lee2004}. In the case of logistic random intercept models, marginal coefficients $\pmb\beta$ can be obtained based on conditional parameters $\pmb\gamma$ as follows:
	\begin{equation}\label{marginal_coefficients}
		\pmb\beta=\dfrac{\pmb\gamma}{\sqrt{1+c^2\sigma_u^2}},
	\end{equation}
	where $c=(16\sqrt{3})/(15\pi)$ \citep{Diggle2002}. Let us denote as $\hat{\pmb\beta}_{\text{M3}}$ the coefficient estimates obtained based on $\hat{\pmb\gamma}$ and $\hat\sigma_{u}^2$.

\end{itemize}

The goal is to analyze the performance of the above-mentioned methods by comparing the estimates $\hat{\pmb\beta}_{\text{M1}}$, $\hat{\pmb\beta}_{\text{M2}}$ and $\hat{\pmb\beta}_{\text{M3}}$ to the true finite population coefficients $\pmb\beta^{True}$.

\section{Simulation study}\label{simulationstudy}

In this section, we describe the simulation study that we have conducted in order to analyze the behavior of the estimation methods described in Section \ref{methods} for estimating the coefficients of the logistic regression model based on complex survey data under different scenarios. As mentioned previously, our goal in this study is to compare the coefficient estimates to the true finite population coefficients in real data-based scenarios.

In Section \ref{scenarios} the simulation process is described in detail and in Section \ref{results} the results obtained in the simulation study are shown.

\subsection{Scenarios and set up}\label{scenarios}

In this section, we describe the different scenarios where the simulation study has been conducted and the steps we have followed. The simulation process is described below, step by step: 

\begin{itemize}
	\item[\textbf{Step 1.}] Generate the pseudo-population $U$ of $N$ units from the set of random variables $(Y, \pmb X)$ (see Appendix \ref{pseudopopulation}): $\left\{(y_i, \pmb x_i)\right\}_{i=1}^N$.
	\item[\textbf{Step 2.}] Compute $\pmb \beta^{True}$ by maximizing the population likelihood in \eqref{population_likelihood}.
\end{itemize}
For $r=1,\ldots,R$ repeat the following steps:
\begin{itemize}
	\item[\textbf{Step 3.}] Obtain a sample $S^r\subset U$ by one-step stratified sampling and assign the corresponding sampling weights $w_i, \forall i\in S^r$ (see Appendix \ref{sampling}).
	\item[\textbf{Step 4.}] Fit the models to $S^r$ by the likelihood functions in  \eqref{likelihood_M1}, \eqref{likelihood_M2} and \eqref{likelihood_M3} and obtain $\hat{\pmb \beta}_{\text{M1}}^{r}$, $\hat{\pmb \beta}_{\text{M2}}^{r}$ and $\hat{\pmb \beta}_{\text{M3}}^{r}$, respectively.
\end{itemize}

Finally, for the results obtained based on samples $r=1,\ldots,R$ and for each method $\forall m\in\{\text{M1},\text{M2},\text{M3}\}$, let us define the bias of the coefficient vector estimates as follows:
\begin{equation}\label{bias_r}
\text{bias}^r_d=\hat{\beta}^{r}_{d,m}-\beta^{True}_d,\quad \forall d=0,1,\ldots,p.
\end{equation}
Then, the average bias (AvBias) and the mean squared error (MSE) across $\forall r=1,\ldots,R$ are defined in equations \eqref{bias} and \eqref{MSE}, respectively:
\begin{equation}\label{bias}
\text{AvBias}_d=\dfrac{1}{R}\sum_{r=1}^{R}\left(\text{bias}^r_d\right)=\dfrac{1}{R}\sum_{r=1}^{R}\left(\hat{\beta}^{r}_{d,m}-\beta^{True}_d\right),\quad \forall d=0,1,\ldots,p,
\end{equation}
\begin{equation}\label{MSE}
\text{MSE}_d=\dfrac{1}{R}\sum_{r=1}^{R}\left(\text{bias}^r_d\right)^2=\dfrac{1}{R}\sum_{r=1}^{R}\left(\hat{\beta}^{r}_{d,m}-\beta^{True}_d\right)^2,\quad \forall d=0,1,\ldots,p.
\end{equation}

Two scenarios have been defined based on the two real surveys described in Section \ref{datasets}, ESIE (Scenario 1, hereinafter) and PRA (Scenario 2, hereinafter). One finite pseudo-population was generated based on each of the surveys (described in \textbf{Step 1.}, see Appendix \ref{pseudopopulation}). Those populations were sampled based on the complex sampling designs that were applied by Eustat in the corresponding real surveys (defined in \textbf{Step 3.}, see Appendix \ref{sampling}). A total of $R=500$ samples were obtained from each pseudo-population. 

In addition, two different models were fitted to the finite population as well as to the samples for each of the surveys with different number of covariates (\textbf{Step 2}). In particular, in Scenario 1 models with $p=1$ ($X_1$) and $p=3$ ($X_1$, $X_2$ and $X_3$) covariates were fitted. In the same way, in Scenario 2, the models were fitted with $p=1$ ($X_1$) and $p=4$ ($X_1$, $X_2$, $X_3$ and $X_4$) covariates. 

It should be noted that all the covariates are categorical and one coefficient was estimated for each category, except for the one considered as reference category. In particular, in Scenario 1, a total of $l=7$ parameters (including the intercept, $\beta_0$) are estimated for the model with $p=1$ covariates and $l=14$ parameters for $p=3$. In the same way, in Scenario 2, $l=7$ parameters are estimated for $p=1$ and $l=14$ parameters for $p=4$.

All computations were performed in (64 bit) R 4.0.5 (R Core Team, 2021) and a workstation equipped with 32GB of RAM, an Intel i7-8700 processor (3.20 Ghz) and Windows 10 operating system.  In particular, the unweighted logistic regression models (M1) were fitted by means of the \texttt{glm} function from the \texttt{stats} package, the weighted logistic regression models (M2) by means of the \texttt{svyglm} from the \texttt{survey} package \citep{Rsurvey} and the unweighted mixed models with random intercept (M3) by the \texttt{glmer} of the \texttt{lme4} package \citep{Bates2015}.

\subsection{Results}\label{results}

In this section, we describe the results we obtained in both scenarios: Scenario 1 (which is based on the ESIE survey) and Scenario 2 (which is based on the PRA survey). As explained in Section \ref{scenarios}, in each scenario two models were fitted with different number of covariates. Our goal is to compare the estimates obtained based on the three coefficient estimation methods described in Section \ref{methods} (which are the unweighted logistic regression (M1), the weighted logistic regression (M2) and the unweighted logistic regression with random intercept (M3)) to the true finite population coefficients ($\pmb\beta^{True}$), in terms of bias and MSE.

Due to the large number of results obtained, we begin by resuming the main findings. When comparing the performance of the three methods in each scenario, we observe that the results differ depending on the scenario. In Scenario 1, M2 outperforms M1 and M3 in terms of bias and MSE, while the estimates obtained with M2 had a greater variance than the estimates obtained with M1 or M3. On the other hand, in Scenario 2, there are no differences among the results obtained with the three methods. The results also show that the method M2 performs correctly in both scenarios and the results are quite similar in terms of bias (which is negligible in all scenarios) and MSE. However, the performance of M1 and M3 methods in terms of bias (and consequently, also in terms of MSE) differ depending on the scenario, being much lower in Scenario 2 than in Scenario 1. We proceed below to analyze the graphical and numerical results related to each scenario. 

\begin{figure}
		\centering
		\subfigure[]{\includegraphics[width=120mm]{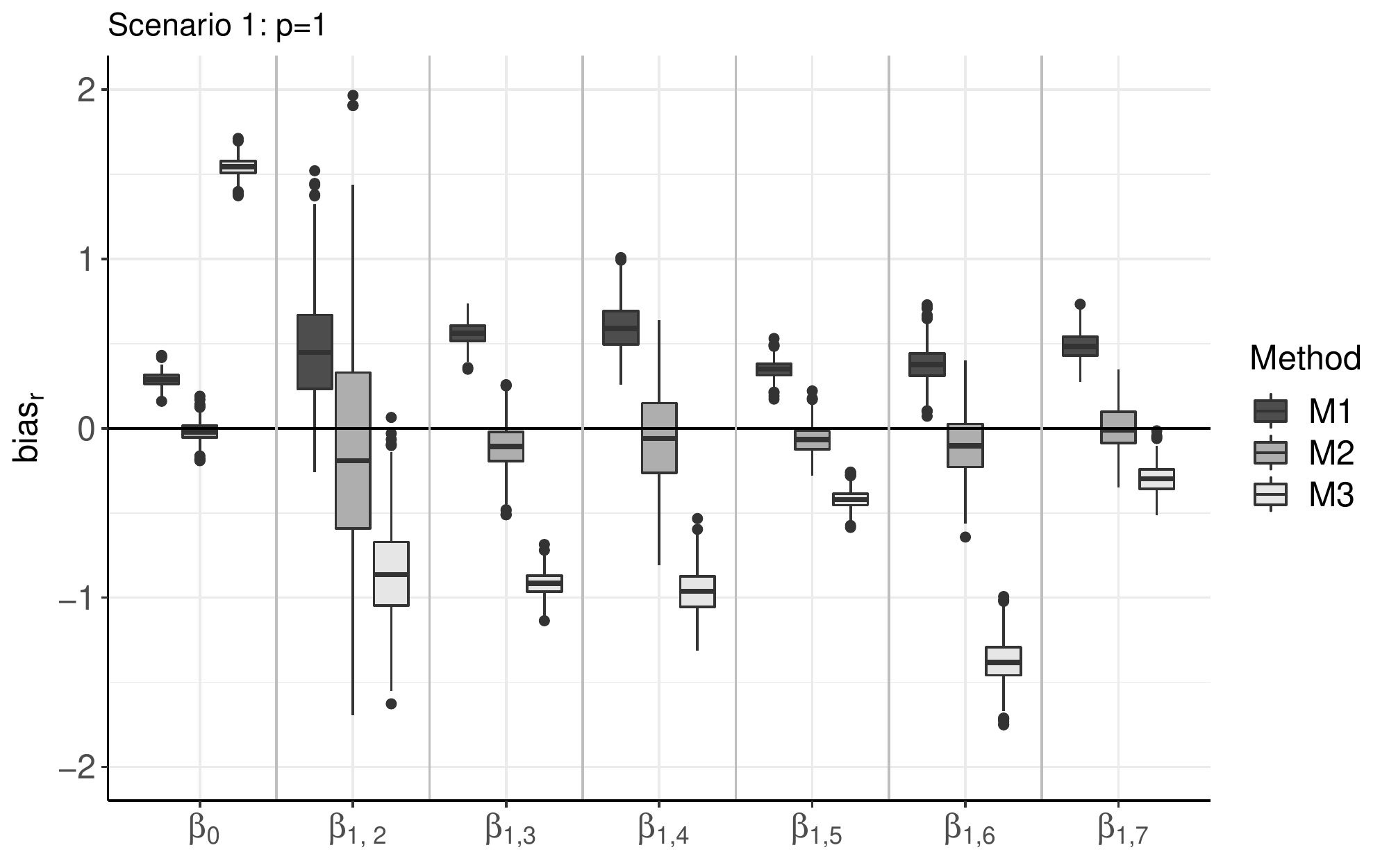}}
		\subfigure[]{\includegraphics[width=120mm]{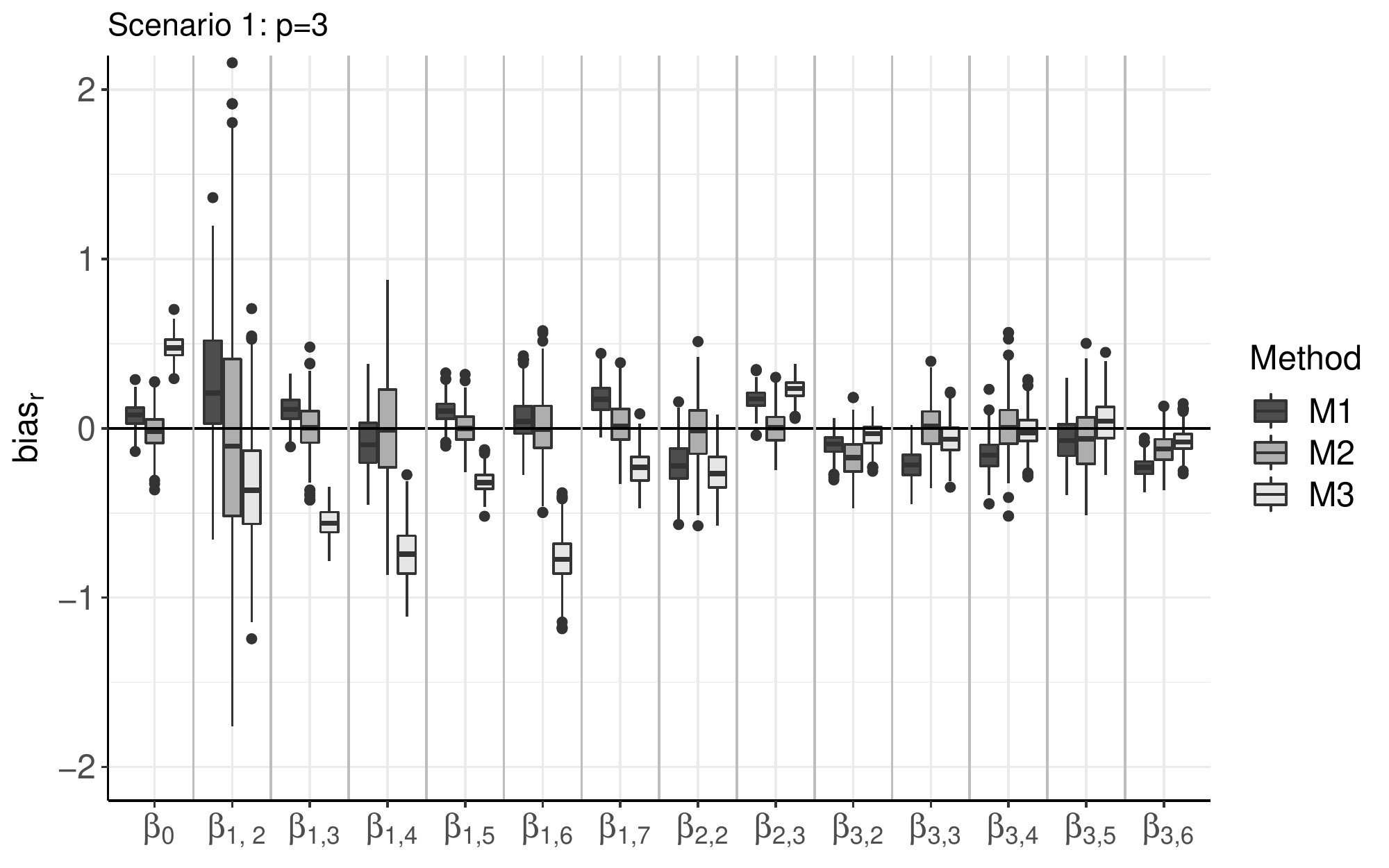}}
		\caption{\small{Box-plots of the bias of the estimates obtained by the methods M1, M2 and M3 for the coefficients in the models with (a) $p=1\:(l=7)$ and (b) $p=3\:(l=14)$ covariates in Scenario 1, $\forall r=1,\ldots,R$.}}
		\label{figesie}
	\end{figure}

Figure \ref{figesie} depicts the box-plots of the bias of the estimates obtained by the methods M1, M2 and M3 for the models with $p=1$ (Figure \ref{figesie} (a)) and $p=3$ (Figure \ref{figesie} (b)) covariates in Scenario 1. As can be observed, M2 is the method that performs the best in terms of bias in both models, with either $p=1$ or $p=3$ covariates. This can also be observed in Table \ref{numerical_results:esie}. This table describes the numerical results of the mean, standard deviation, average bias and MSE of those estimates, as well as the true finite population coefficients in Scenario 1 for the models with $p=1$ and $p=3$ covariates, respectively. As can be seen, while the estimates obtained by M2 method are quite similar to the true coefficients ($\pmb\beta^{True}$) obtained in the finite population (which leads to low average biases for this method), the estimates obtained by M1 and M3 methods differ considerably. In the estimates obtained for the model with $p=1$ for example, for the coefficient $\beta_{1,6}$ for instance, the average bias obtained by means of M2 method is of -0.095, which is considerably lower than the one of the M1 method (0.378) and the M3 method (-1.379). It can also be observed that the average bias decreases for all the methods (and most notably for M1 and M3) when $p=3$ covariates are included into the model. This is in line with \cite{Nathan1980}. In particular, the average bias of the coefficient estimates related to the category $\beta_{1,6}$ decreases to 0.050 for the M1 method, to 0.007 for M2 and to -0.700 for M3 in the model with $p=3$ covariates. 

In Figure \ref{figesie} it can also be seen that the variability of the estimates obtained based on the method M2 is the greatest one, comparing to the rest of the methods. This is also shown in Table \ref{numerical_results:esie}, where the standard deviations of these estimates can be up to twice as large as that of M1 and M3. For example, the standard deviations corresponding to the estimates of $\beta_{1,3}$ are 0.063, 0.132 and 0.070 for M1, M2 and M3, respectively. The larger variability of the weighted estimates has also been observed in previous studies (see, e.g., \cite{Scott1986}). The source of variability could also be related to data. It is especially remarkable the variability of the estimates of the coefficient $\beta_{1,2}$ for all the methods in general, and most importantly for M2. It should be noted that there are very few units in the category 2 of the covariate $X_1$ in Scenario 1. In particular, 450  units in the population (0.2\% of the total of units in the finite population) take this category on that covariate and in the samples this amount varies from 27 (0.3\%) to 53 (0.7\%) (results not shown). This may be affecting in the estimates of the parameter $\beta_{1,2}$, specifically for the M2 method. The behaviour of the estimates of $\beta_{1,4}$ could be explained in the same way, for which a greater variability is also observed, especially for M2 (2008 units (1.0\%) in the finite population, from 178 (2.5\%) to 232 (3.2\%) in the samples). In addition, in Table \ref{numerical_results:esie}, it should also be noted that for all the methods, the standard deviation of the three methods are slightly greater for the model with $p=3$ than for the one with $p=1$ covariates. 

Finally, as shown in Table \ref{numerical_results:esie}, the method M3 is, in most of the cases, the one with the greatest MSE, because of the large bias of the estimates based on that method. For instance, the MSE of the coefficient corresponding to the category $\beta_{1,4}$ in the model with $p=1$ is 0.722 for the method M3, while for the M2 and M1 methods the MSE are 0.085 and 0.378, respectively. Given that the bias decreases while adding covariates for the methods M1 and M3, the MSE also decreases in the same way. For the same coefficient,  when $p=3$, the MSE related to the method M3 decreases to 0.536. The MSE of the M2 method is quite similar in both models, with $p=1$ and $p=3$ covariates. Comparing the MSE of M2 and M1 methods it can be observed that the MSE of M1 is greater when $p=1$. However, in Scenario 1 with $p=3$, there are no differences in terms of MSE between M1 and M2 due to the larger variability of M2 estimates despite their smaller bias. 

\begin{figure}
		\centering
		\subfigure[]{\includegraphics[width=120mm]{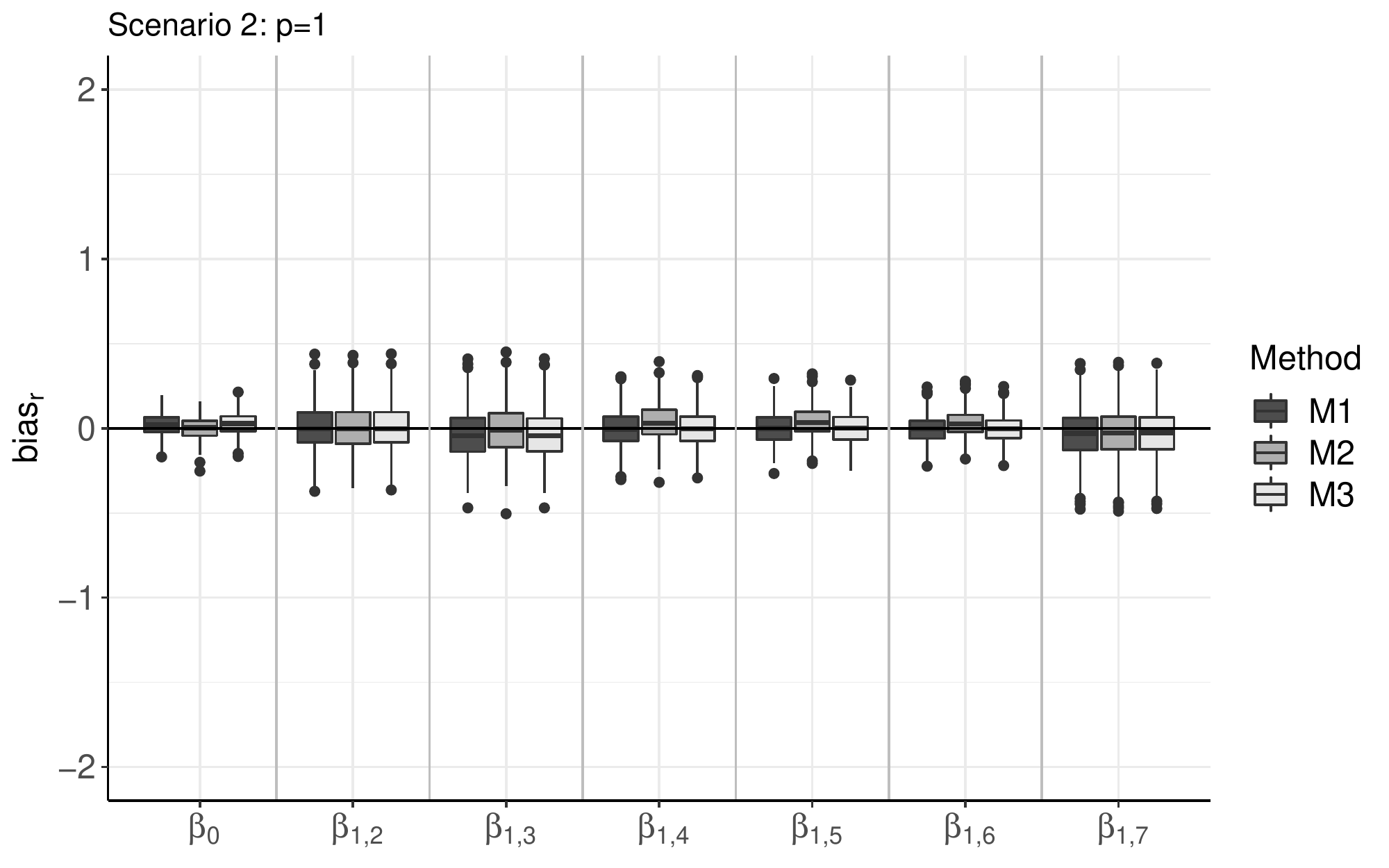}}
		\subfigure[]{\includegraphics[width=120mm]{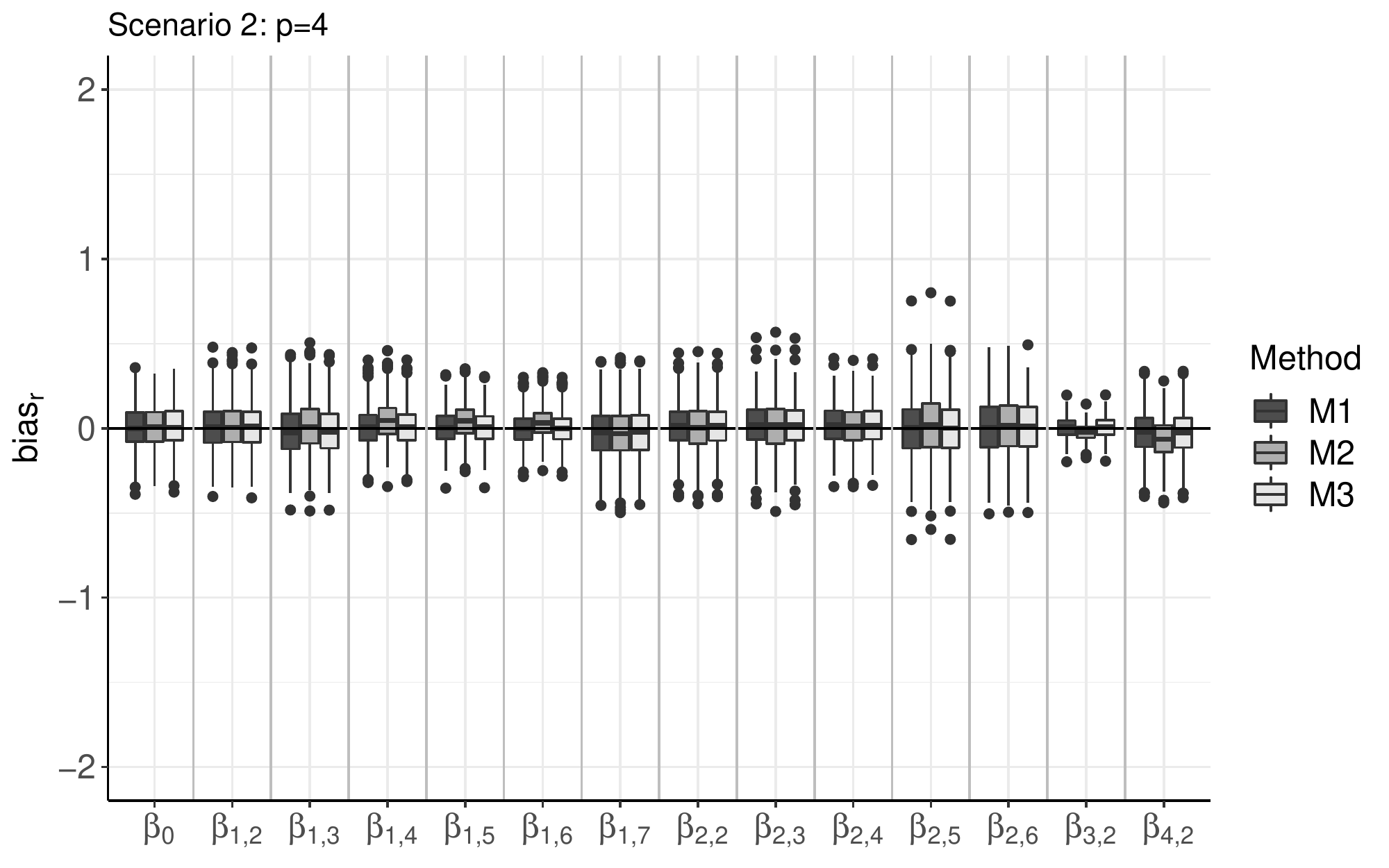}}
		\caption{\small{Box-plots of the bias of the estimates obtained by the methods M1, M2 and M3 for the coefficients in the models with (a) $p=1\:(l=7)$ and (b) $p=4\:(l=14)$ covariates in Scenario 2, $\forall r=1,\ldots,R$.}}
		\label{figpra}
	\end{figure}

Figure \ref{figpra} depicts the box-plots of the bias of the estimates obtained by the methods M1, M2 and M3 for the models with $p=1$ and $p=4$ covariates in Scenario 2. In this case, as shown in Figure \ref{figpra}, the performance of the three methods is quite similar in terms of bias and variability. The differences are not considerable, neither among the different methods, nor between the different models (fitted with $p=1$ and $p=4$ covariates). Table \ref{numerical_results:pra} describes the numerical results of the mean, standard deviation, average bias and MSE of those estimates and the true finite population coefficients for $p=1$ and $p=4$ in Scenario 2. The average bias is very low for all the methods and in both models, either with $p=1$ or $p=4$ covariates. The greatest observed average bias is -0.067, which corresponds to the coefficient $\beta_{4,2}$ of the model with $p=4$ covariates for the method M2. The variability of the estimates obtained by the method M2 are usually slightly greater than that of the rest of the methods. However, as noted above, those differences are very small. The greatest difference in terms of standard deviation of the estimates and MSE are observed in the model with $p=4$ for the coefficient estimates corresponding to category $\beta_{2,5}$. The standard deviation of the estimates obtained by means of M2 method is 0.185 while the ones corresponding to the M1 and M3 methods are 0.174. In the same way, the MSE of the M2 method for this coefficient is 0.035 while for the methods M1 and M3 is 0.030. It can be concluded that all the studied methods perform properly to estimate the finite population model coefficients in Scenario 2. 

\begin{landscape}
	
	\begin{table}[h]
		\centering
		\caption{True finite population model coefficients ($\pmb \beta^{True}$) and the average (mean), standard deviation (sd), average bias (AvBias) and MSE of the estimates obtained by the M1, M2 and M3 methods for the models with $p=1$ and $p=3$ covariates in Scenario 1 for $R=500$ samples.}
		\begin{tabular}{r|r|rrr|rrr|rrr}
			\hline
			&   \multicolumn{1}{c}{$\pmb \beta^{True}$}    & \multicolumn{3}{|c}{M1} 		& \multicolumn{3}{|c}{M2}       & \multicolumn{3}{|c}{M3}        \\
			\hline
			&  & Mean(sd) & AvBias & MSE      & Mean(sd) & AvBias & MSE      & Mean(sd) & AvBias & MSE\\
			\hline
			\multicolumn{11}{l}{$\mathbf{p=1\:(l=7)}$} \\
			\hline
			$\beta_0$ & -1.015 & -0.727(0.043) & 0.288 & 0.085 & -1.035(0.058) & -0.021 & 0.004 & 0.529(0.056) & 1.544 & 2.387\\
			$\beta_{1,2}$ & 1.184 & 1.658(0.325) & 0.474 & 0.330 & 1.050(0.659) & -0.134 & 0.451 & 0.334(0.290) & -0.850 & 0.807\\
			$\beta_{1,3}$ & 1.360 & 1.920(0.063) & 0.560 & 0.318 & 1.252(0.132) & -0.108 & 0.029 & 0.443(0.070) & -0.916 & 0.845\\
			$\beta_{1,4}$ & 1.342 & 1.942(0.138) & 0.600 & 0.378 & 1.283(0.286) & -0.060 & 0.085 & 0.382(0.141) & -0.960 & 0.942\\
			$\beta_{1,5}$ & 0.537 & 0.885(0.055) & 0.348 & 0.124 & 0.471(0.085) & -0.067 & 0.012 & 0.118(0.053) & -0.420 & 0.179\\
			$\beta_{1,6}$ & 1.908 & 2.286(0.110) & 0.378 & 0.155 & 1.813(0.184) & -0.095 & 0.043 & 0.531(0.129) & -1.377 & 1.912\\
			$\beta_{1,7}$ & 0.470 & 0.956(0.080) & 0.487 & 0.243 & 0.471(0.121) & 0.001 & 0.015 & 0.174(0.088) & -0.296 & 0.095\\	
			\hline
			\multicolumn{11}{l}{$\mathbf{p=3\:(l=14)}$} \\
			\hline
			$\beta_0$ & -0.959 & -0.883(0.069) & 0.076 & 0.010 & -0.979(0.111) & -0.020 & 0.013 & -0.484(0.066) & 0.475 & 0.230 \\
			$\beta_{1,2}$ & 0.602 & 0.850(0.356) & 0.248 & 0.188 & 0.582(0.668) & -0.020 & 0.445 & 0.257(0.330) & -0.344 & 0.227 \\
			$\beta_{1,3}$ & 0.824 & 0.937(0.078) & 0.112 & 0.019 & 0.830(0.146) & 0.006 & 0.021 & 0.269(0.082) & -0.556 & 0.316 \\
			$\beta_{1,4}$ & 0.926 & 0.840(0.158) & -0.086 & 0.032 & 0.926(0.310) & 0.000 & 0.096 & 0.184(0.161) & -0.742 & 0.576 \\
			$\beta_{1,5}$ & 0.382 & 0.482(0.066) & 0.101 & 0.014 & 0.383(0.096) & 0.002 & 0.009 & 0.065(0.064) & -0.317 & 0.104 \\
			$\beta_{1,6}$ & 1.145 & 1.195(0.124) & 0.050 & 0.018 & 1.153(0.197) & 0.007 & 0.039 & 0.374(0.138) & -0.771 & 0.614 \\
			$\beta_{1,7}$ & 0.355 & 0.526(0.089) & 0.172 & 0.037 & 0.373(0.125) & 0.018 & 0.016 & 0.119(0.097) & -0.235 & 0.065 \\
			$\beta_{2,2}$   & -0.630 & -0.843(0.136) & -0.212 & 0.064 & -0.647(0.182) & -0.016 & 0.033 & -0.892(0.121) & -0.262 & 0.083 \\
			$\beta_{2,3}$   & 0.036 & 0.207(0.058) & 0.170 & 0.032 & 0.040(0.098) & 0.004 & 0.010 & 0.266(0.056) & 0.230 & 0.056 \\
			$\beta_{3,2}$ & 0.042 & -0.055(0.065) & -0.097 & 0.014 & -0.128(0.115) & -0.171 & 0.042 & 0.005(0.067) & -0.038 & 0.006 \\
			$\beta_{3,3}$ & 1.514 & 1.297(0.086) & -0.217 & 0.054 & 1.520(0.143) & 0.006 & 0.020 & 1.448(0.091) & -0.066 & 0.013 \\
			$\beta_{3,4}$ & 1.418 & 1.260(0.096) & -0.158 & 0.034 & 1.428(0.150) & 0.010 & 0.023 & 1.402(0.098) & -0.016 & 0.010 \\
			$\beta_{3,5}$ & 1.564 & 1.491(0.134) & -0.073 & 0.023 & 1.497(0.190) & -0.067 & 0.040 & 1.601(0.133) & 0.037 & 0.019 \\
			$\beta_{3,6}$ & 2.700 & 2.465(0.053) & -0.234 & 0.058 & 2.577(0.093) & -0.123 & 0.024 & 2.623(0.069) & -0.077 & 0.011 \\
			\hline
		\end{tabular}%
	\label{numerical_results:esie}
	\end{table}%
	
\end{landscape}

\begin{landscape}
	\begin{table}[htbp]
		\centering
		\caption{True finite population model coefficients ($\pmb \beta^{True}$) and the average (mean), standard deviation (sd), average bias (AvBias) and MSE of the estimates obtained by the M1, M2 and M3 methods for the models with $p=1$ and $p=4$ covariates in Scenario 2 for $R=500$ samples.}
		\begin{tabular}{r|r|rrr|rrr|rrr}
			\hline
			&   \multicolumn{1}{c}{$\pmb \beta^{True}$}    & \multicolumn{3}{|c}{M1} 		& \multicolumn{3}{|c}{M2}       & \multicolumn{3}{|c}{M3}        \\
			\hline
			&  & Mean(sd) & AvBias & MSE   & Mean(sd) & AvBias & MSE  & Mean(sd) & AvBias & MSE \\
			\hline
			\multicolumn{11}{l}{$\mathbf{p=1\:(l=7)}$} \\
			\hline
			$\beta_0$ & -0.965 & -0.945(0.068) & 0.020 & 0.005 & -0.965(0.065) & 0.000 & 0.004 & -0.941(0.068) & 0.025 & 0.005 \\
			$\beta_{1,2}$ & 2.676 & 2.681(0.127) & 0.005 & 0.016 & 2.680(0.133) & 0.003 & 0.018 & 2.682(0.127) & 0.006 & 0.016 \\
			$\beta_{1,3}$ & 3.369 & 3.337(0.149) & -0.032 & 0.023 & 3.366(0.152) & -0.003 & 0.023 & 3.338(0.149) & -0.031 & 0.023 \\
			$\beta_{1,4}$ & 3.539 & 3.539(0.110) & 0.000 & 0.012 & 3.576(0.109) & 0.038 & 0.013 & 3.539(0.109) & 0.001 & 0.012 \\
			$\beta_{1,5}$ & 3.034 & 3.035(0.093) & 0.002 & 0.009 & 3.071(0.091) & 0.037 & 0.010 & 3.036(0.093) & 0.003 & 0.009 \\
			$\beta_{1,6}$ & 1.571 & 1.568(0.082) & -0.003 & 0.007 & 1.600(0.075) & 0.029 & 0.006 & 1.569(0.082) & -0.002 & 0.007 \\
			$\beta_{1,7}$ & -2.854 & -2.886(0.139) & -0.033 & 0.020 & -2.883(0.146) & -0.029 & 0.022 & -2.884(0.139) & -0.030 & 0.020 \\
			\hline
			\multicolumn{11}{l}{$\mathbf{p=4\:(l=14)}$} \\
			\hline
			$\beta_0$ & -1.529 & -1.523(0.125) & 0.006 & 0.016 & -1.524(0.127) & 0.005 & 0.016 & -1.516(0.125) & 0.012 & 0.016 \\
			$\beta_{1,2}$ & 2.478 & 2.490(0.134) & 0.012 & 0.018 & 2.490(0.140) & 0.012 & 0.020 & 2.490(0.134) & 0.012 & 0.018 \\
			$\beta_{1,3}$ & 3.206 & 3.188(0.153) & -0.017 & 0.024 & 3.221(0.156) & 0.015 & 0.024 & 3.189(0.153) & -0.017 & 0.024 \\
			$\beta_{1,4}$ & 3.329 & 3.342(0.121) & 0.013 & 0.015 & 3.379(0.122) & 0.050 & 0.017 & 3.341(0.121) & 0.013 & 0.015 \\
			$\beta_{1,5}$ & 2.780 & 2.785(0.103) & 0.005 & 0.011 & 2.824(0.103) & 0.044 & 0.012 & 2.786(0.103) & 0.006 & 0.011 \\
			$\beta_{1,6}$ & 1.379 & 1.376(0.097) & -0.003 & 0.009 & 1.413(0.092) & 0.034 & 0.010 & 1.378(0.097) & -0.001 & 0.009 \\
			$\beta_{1,7}$ & -2.943 & -2.974(0.149) & -0.031 & 0.023 & -2.974(0.157) & -0.031 & 0.026 & -2.970(0.150) & -0.028 & 0.023 \\
			$\beta_{2,2}$ & 0.702 & 0.715(0.134) & 0.013 & 0.018 & 0.704(0.144) & 0.002 & 0.021 & 0.714(0.134) & 0.012 & 0.018 \\
			$\beta_{2,3}$ & 1.391 & 1.407(0.140) & 0.016 & 0.020 & 1.404(0.148) & 0.013 & 0.022 & 1.406(0.140) & 0.015 & 0.020 \\
			$\beta_{2,4}$ & 0.811 & 0.827(0.117) & 0.016 & 0.014 & 0.819(0.126) & 0.008 & 0.016 & 0.827(0.117) & 0.016 & 0.014 \\
			$\beta_{2,5}$ & 1.620 & 1.618(0.174) & -0.001 & 0.030  & 1.638(0.185) & 0.018 & 0.035 & 1.618(0.174) & -0.001 & 0.030 \\
			$\beta_{2,6}$ & 1.664 & 1.669(0.159) & 0.005 & 0.025 & 1.678(0.165) & 0.014 & 0.027 & 1.670(0.159) & 0.006 & 0.025 \\
			$\beta_{3,2}$ & -0.427 & -0.422(0.057) & 0.005 & 0.003 & -0.451(0.049) & -0.024 & 0.003 & -0.421(0.057) & 0.005 & 0.003 \\
			$\beta_{4,2}$ & -0.481 & -0.506(0.129) & -0.025 & 0.017 & -0.548(0.123) & -0.067 & 0.019 & -0.508(0.129) & -0.027 & 0.017 \\
			\hline
		\end{tabular}%
		\label{numerical_results:pra}
	\end{table}%
\end{landscape}

\section{Application to the real data sets}\label{application}

In this section we apply the methods described in Section \ref{methods} to the real survey data described in Section \ref{datasets}. The goal is to compare the coefficient estimates obtained by means of the different methods among them. Note that in this case, the real finite population coefficients are not known.

One model was fitted to each of the surveys. In particular, we fitted the model with three covariates ($p=3$) to the ESIE survey and the model with four covariates ($p=4$) to the PRA survey. Those covariates are the ones that were considered in the simulation study for both surveys and are also considered in the models that are applied in practice by Eustat. To fit those models, the three methods described in Section \ref{methods} were applied: the unweighted logistic regression (M1), the weighted logistic regression (M2) and the unweighted logistic regression with random intercept (M3). Table \ref{application_esie} and Table \ref{application_pra} depict the coefficient estimates and their standard errors obtained for models fitted to the ESIE and PRA surveys respectively.
 
\begin{table}[h]
	\centering
	\footnotesize
	\caption{Coefficient estimates (Estimate) and their standard errors (SE) obtained by means of the methods M1, M2 and M3 for the ESIE survey with $p=3$ covariates.}
	\begin{tabular}{r|rr|rr|rr}
		\hline
		\multicolumn{7}{c}{ESIE survey}     \\  
		\hline                                 
		\multirow{2}{*}{} & \multicolumn{2}{c|}{M1} & \multicolumn{2}{c|}{M2} & \multicolumn{2}{c}{M3} \\ \cline{2-7}
		& Estimate    & SE       & Estimate    & SE       & Estimate    & SE       \\
		\hline
		$\beta_0$ & -2.261      & 0.097    & -2.482      & 0.133    & -2.217      & 0.140    \\
		$\beta_{1,2}$ & 1.892       & 0.338    & 1.293       & 0.444    & 1.697       & 0.368    \\
		$\beta_{1,3}$ & 2.490       & 0.107    & 2.718       & 0.161    & 2.337       & 0.119    \\
		$\beta_{1,4}$ & 2.248       & 0.196    & 2.577       & 0.299    & 2.151       & 0.215    \\
		$\beta_{1,5}$ & 1.550       & 0.084    & 1.721       & 0.111    & 1.458       & 0.094    \\
		$\beta_{1,6}$ & 2.260       & 0.146    & 2.544       & 0.206    & 2.092       & 0.181    \\
		$\beta_{1,7}$ & 1.341       & 0.103    & 1.130       & 0.133    & 1.197       & 0.119    \\
		$\beta_{2,2}$ & -0.774      & 0.148    & -0.613      & 0.189    & -0.883      & 0.329    \\
		$\beta_{2,3}$ & 0.453       & 0.073    & 0.358       & 0.107    & 0.538       & 0.123    \\
		$\beta_{3,2}$ & 0.669       & 0.069    & 0.632       & 0.097    & 0.750       & 0.077    \\
		$\beta_{3,3}$ & 0.996       & 0.096    & 0.965       & 0.132    & 1.124       & 0.134    \\
		$\beta_{3,4}$ & 1.479       & 0.114    & 1.452       & 0.152    & 1.698       & 0.149    \\
		$\beta_{3,5}$ & 2.230       & 0.182    & 2.205       & 0.241    & 2.461       & 0.209    \\
		$\beta_{3,6}$ & 2.454       & 0.143    & 2.532       & 0.151    & 2.787       & 0.195   \\
		\hline
	\end{tabular}
	\label{application_esie}
\end{table}

\begin{table}[h]
	\centering
	\footnotesize
	\caption{Coefficient estimates (Estimate) and their standard errors (SE) obtained by means of the methods M1, M2 and M3 for the PRA survey with $p=4$ covariates.}
	\begin{tabular}{r|rr|rr|rr}
		\hline
		\multicolumn{7}{c}{PRA survey} \\	
		\hline	
		\multirow{2}{*}{} & \multicolumn{2}{c|}{M1} & \multicolumn{2}{c|}{M2} & \multicolumn{2}{c}{M3} \\ \cline{2-7}
		& Estimate    & SE       & Estimate    & SE       & Estimate    & SE       \\
		\hline
		$\beta_0$ & -2.039      & 0.176    & -2.040      & 0.171    & -2.037      & 0.179    \\
		$\beta_{1,2}$ & 2.508       & 0.164    & 2.523       & 0.172    & 2.515       & 0.164    \\
		$\beta_{1,3}$ & 3.106       & 0.179    & 3.105       & 0.191    & 3.113       & 0.179    \\
		$\beta_{1,4}$ & 3.191       & 0.121    & 3.292       & 0.126    & 3.194       & 0.122    \\
		$\beta_{1,5}$ & 2.836       & 0.114    & 2.934       & 0.118    & 2.835       & 0.114    \\
		$\beta_{1,6}$ & 1.455       & 0.103    & 1.543       & 0.108    & 1.454       & 0.103    \\
		$\beta_{1,7}$ & -3.170      & 0.184    & -3.102      & 0.199    & -3.182      & 0.184    \\
		$\beta_{2,2}$ & 1.005       & 0.174    & 0.899       & 0.177    & 1.016       & 0.174    \\
		$\beta_{2,3}$ & 1.689       & 0.178    & 1.587       & 0.182    & 1.700       & 0.179    \\
		$\beta_{2,4}$ & 1.167       & 0.171    & 1.056       & 0.170    & 1.170       & 0.172    \\
		$\beta_{2,5}$ & 2.123       & 0.207    & 1.970       & 0.227    & 2.128       & 0.208    \\
		$\beta_{2,6}$ & 2.357       & 0.192    & 2.177       & 0.201    & 2.360       & 0.193    \\
		$\beta_{3,2}$ & -0.596      & 0.063    & -0.546      & 0.067    & -0.596      & 0.063    \\
		$\beta_{4,2}$ & 0.547       & 0.158    & 0.530       & 0.190    & 0.551       & 0.159   \\
		\hline
	\end{tabular}
	\label{application_pra}
\end{table}

As shown in Table \ref{application_esie}, the coefficient estimates, as well as their standard errors, obtained by means of the three above-mentioned methods differ considerably in the ESIE survey. It should be noted that these differences in the estimations and their standard errors, could lead to considerable differences in the Wald statistic defined as the fraction among those parameters. However, in this case, those differences did not affect the significance of the model parameters and all of them are statistically significant (results not shown). The largest standard errors are in most of the cases the ones obtained by means of the method M2. In addition, the standard errors related to the coefficient $\beta_{1,2}$  are larger than any other's, which is in line with the large variability observed in the simulation study for this coefficient (in Scenario 1). Based on the results obtained in the simulation study, we may conclude that the model fitted by the method M2 would be the preferred one in this case.

In contrast, the coefficient estimates and their standard errors obtained for the PRA survey are very similar among them, as can be observed in Table \ref{application_pra}. This is also in line with the results observed in the simulation study (in Scenario 2). As expected, the standard errors of the estimates obtained by M2 are usually slightly greater than the rest, although there are not great differences, in general. 

\section{Discussion}\label{discussion}

In this work we compared the performance of three different methods to estimate model coefficients in the logistic regression framework for complex survey data by means of a real data based simulation study. In general, the results we obtained are in line with the ones obtained in related works, based on either logistic \citep{Scott1986, Scott2002, Lumley2017a, Chambless1985, Reiter2005} or linear regression framework \citep{DeMets1977, Holt1980, Nathan1980, Smith1981}. Nevertheless, there are also some differences among this work and the above-mentioned studies. We proceed to comment on these similarities and differences in the following lines.

One of the greatest differences between this study and the ones mentioned previously is that this work is a simulation study based on real survey data. The objective has been to work in a realistic scenario that allows us to compare the results we obtain to the true coefficients of the finite population models. Data for the simulation study have been simulated based on two real surveys conducted by the Official Statistics Basque Office (Eustat). In both surveys the finite population were sampled by one-step stratification. However, the strata were defined in very different ways. In the ESIE survey the strata were defined by means of the combination of three categorical variables with many categories, resulting in a total of $585$ small strata. On the other hand, in the PRA survey, strata were defined by means of the region to which each individual belongs, which leads to 23 different strata. In addition to the sampling design, the impact of the number of covariates included in the model and the number of parameters, were also analyzed. It should be noted that in this simulation study the theoretical model from which the finite population is generated from is not known for us. Thus, we compare the model estimates obtained based on the methods under study to the true coefficient values obtained by fitting the model to the finite population. 

The main conclusions of this study are that the weighted logistic regression (M2) performed properly in both scenarios and the estimates we obtained were unbiased. In contrast, the behavior of the unweighted logistic regression (M1) and the unweighted logistic regression with random intercept (M3) depended on the scenario and on the number of covariates/parameters estimated in the model. In the scenario related to the ESIE survey, unlike in the scenario based on the PRA survey, biased estimates were obtained based on these two methods. These results are in line with \cite{Scott1986, Holt1980, Nathan1980} among others, which also warn about the bias of the unweighted coefficient estimates in both, linear and logistic regression frameworks. \cite{Scott1986} claim that the bias of the unweighted coefficient estimates is smaller when the model fitted to the sample is exactly the same as the true theoretical model from which the data is derived than when the model fitted is ``reasonable but not perfect''. As mentioned previously, the theoretical model from which the finite population is generated from is not known for us. Nevertheless, in this study we have also observed that the bias becomes smaller when more covariates are included into the model, which would be in line with the results obtained in the above-mentioned studies. However, this bias is still larger than the bias obtained by means of the weighted logistic regression.

The variability of the estimates obtained by the weighted logistic regression model is greater than that of the estimates obtained by means of the unweighted logistic regression model (with and/or without random intercept) which is in line with \cite{Chambless1985, Lumley2017a, Scott1986}. These differences are not very large in most of the cases. However, we have observed that when there are few individuals in a particular category of a categorical variable, then the variability of the weighted estimates of the coefficient corresponding to that category can be much greater than the unweighted ones. We conclude that we should be careful when we have categorical variables with unbalanced distribution of individuals in the categories.

We also applied the three methods under study to real survey data and the estimates we obtained are in line with the results observed in the simulation study. On the one hand, in the PRA survey, the estimates are quite similar among them, and there are not many differences between the standard deviations of these estimates, which leads us to conclude that all the studied methods work properly in this case. On the other hand, in the ESIE survey, there are many differences in the estimates of the parameters among different methods. Observing the similarities among the simulation study and the application to real data sets, and taking into account that those results are also in line with the results obtained in similar empirical studies, such as \cite{Chambless1985} and \cite{Lumley2017a}, we can assume that the weighted logistic regression would be preferred when working with ESIE survey data.

We now proceed to comment on the limitations of this work. First of all, in this simulation study we are unable to know which is the theoretical model from which the data is derived due to the fact that we aimed for the simulation study to be based on real survey data and hence, we have focused on comparing the estimates obtained based on the samples with the true coefficients of the model fitted to the finite population. It should be noted that often the objective in working with survey data is to draw conclusions related to that particular finite population, and therefore, this comparative study makes sense in that context. For those readers who are interested in comparisons with the theoretical infinite population model, we suggest checking \cite{Scott2002}. Secondly, as mentioned above, some authors recommend including the design variables and the interactions between them as covariates in the model. However, in this case, and in particular in the case of the ESIE survey, this option would not be feasible due to the large number of parameters (a total of $585$) to be estimated within the model. Therefore, we have decided to fit the mixed model, replicating in this way the comparison made by \cite{Lumley2017a} on real datasets. In addition, some of the covariates included in the models are related to the stratification variables. It should also be noted that in this simulation study we have worked with surveys of considerable sample sizes, which is quite common in official statistics. Nevertheless, we also believe that it would be interesting to work with simulations based on real surveys with smaller sample sizes and compare the results, paying special attention to the variability of the estimates. Lastly, in this work we have focused on the estimation of the parameters of the logistic regression model. Other issues of interest, such as the selection of the covariates or the effect that these differences may have on the estimated probabilities of the individuals, are out of the scope of this work.  

To sum up, the weighted logistic regression performs properly in all the scenarios we have drawn. In contrast, the behavior of the unweighted logistic regression (both, with and without random intercept) depends on the scenario. Therefore, based on the results of the simulation study, we believe that not using sampling weights when necessary leads to worse results than using them when they are not needed. For this reason, we would recommend the use of the weighted logistic regression model in the context of complex survey data.

\noindent {\bf{Acknowledgment}}\\
\noindent This work was financially supported in part by grants from the Departamento de Educaci\'on, Pol\'itica Ling\"u\'istica y Cultura del Gobierno Vasco [IT1456-22] and by the Ministry of Science and Innovation through BCAM Severo Ochoa accreditation [CEX2021-001142-S / MICIN / AEI / 10.13039/501100011033] and through project [PID2020-115882RB-I00 / AEI / 10.13039/501100011033] funded by Agencia Estatal de Investigaci\'on and acronym ``S3M1P4R" and also by the Basque Government through the BERC 2022-2025 program. The work of AI was supported by grant [PIF18/213].

We would like to acknowledge the Official Statistics Basque Office (Eustat) for providing us with the ESIE and PRA survey data.

\vspace*{1pc}

\noindent {\bf{Conflict of interest}}

\noindent {\it{The authors declare that there are no conflicts of interest.}}

\bibliographystyle{chicago}
\bibliography{ComplexSurvey}

\appendix

\section{Pseudo-population generation}\label{pseudopopulation}

This section describes the process of generating the pseudo-populations that have been used in the simulation study described in Section \ref{simulationstudy}, in Scenario 1 (based on the ESIE survey) and in Scenario 2 (based on the PRA survey).  

The pseudo-population applied in Scenario 2, related to the PRA survey, is actually a real finite population, for which the response variable, as well as the rest of the explanatory variables, are known. This pseudo-population was obtained and provided by Eustat.

In the case of Scenario 1, we have generated a pseudo-population based on the real finite population and sample of the ESIE survey. Let us denote as $S_{\text{ESIE}}$ the original survey sample and $U_{\text{ESIE}}$ the real finite population of size $N$ ($S_{\text{ESIE}}\subset U_{\text{ESIE}}$). As explained in Section \ref{datasets}, a total of $H$ strata have been defined (i.e., $\{1,\ldots,H\}$) combining information of three categorical variables, which will be denoted as $X_1$, $X_2$ and $X_3$. Therefore, the finite population can be partitioned in subsets defined by means of these strata, i.e., $U_{\text{ESIE}}=\bigcup_{h=1}^H U_{\text{ESIE},h}$. $\forall h\in\{1,\ldots,H\}$ let us indicate as $N_h$ the size of stratum $h$ in the finite population $U_{\text{ESIE}}$ $(U_{\text{ESIE},h})$ and as $n_h$ the size of this stratum in the sample $S_{\text{ESIE}}$. Then, the sampling weight associated to a unit $j\in S_{\text{ESIE}}$ from stratum $h$ is the following:
\begin{equation}
    w_j=\dfrac{N_h}{n_h}.
\end{equation}

Our goal is to generate a pseudo-population $(U)$ based on the known real ESIE survey data, for which all the information of the covariates $X_1,\ldots,X_p$ and the response variables $Y_1,\ldots,Y_q$ will be available. This new pseudo-population $U$ will be the same size as the true ESIE population ($N$). In order to ease the notation, the variable names of the pseudo-population are the same as in the real finite population and the units of the real ESIE population will be denoted as $j\in U_{\text{ESIE}}$ while the units that are artificially generated for the pseudo-population will be denoted as $i\in U$.

Several dichotomous response variables are available in the original survey being the response variable $Y$, the one we have applied in the simulation study, one of them. All possible combinations of these response variables have been examined. For instance, assuming that $Y_1,\ldots,Y_{q}$ are all the response variables that are available in the survey (where $Y\in\{Y_1,\ldots,Y_{q}\}$), for some $j\in S_{\text{ESIE}}$: $\pmb y_j=(y_{1,j},\ldots,y_{q,j})=\alpha$, $\forall\alpha\in\left\{\alpha_1,\ldots,\alpha_A\right\}$, where $\left\{\alpha_1,\ldots,\alpha_A\right\}$ is the set of all of possible combinations of the responses. For each stratum $h\in\{1,\ldots,H\}$ and for each possible combination of the responses (i.e., $\forall\alpha\in\left\{\alpha_1,\ldots,\alpha_A\right\})$ we generate $N_{h,\alpha}$ units in the pseudo-population ($U$) as:
\begin{equation}
N_{h,\alpha}=\sum_{j\in S_{\text{ESIE}}} w_j 1_{U_{\text{ESIE},\:h}}(j)\left[\pmb y_j=\alpha\right],
\end{equation}
where,
\begin{equation}\label{indicator_function}
	1_{U_{\text{ESIE},\:h}}(j)=\left\{
	\begin{array}{cc}
		1, &\text{if $j\in U_{\text{ESIE},\:h}$},\\
		0, &\text{if $j\notin U_{\text{ESIE},\:h}$},\\
	\end{array}
	\right.
\end{equation}
and
\begin{equation}
\left[\pmb y_j=\alpha\right]=\left\{
\begin{array}{ccc}
1 & \text{if} & \pmb y_j=\alpha, \\
0 & \text{if} & \pmb y_j\neq\alpha. \\
\end{array}
\right.
\end{equation}
In this way, $N_{h,\alpha}$ is the number of units of the pseudo-population $U$ in stratum $h$, which take the values of responses $(y_{1,j},\ldots,y_{q,j})=\alpha$. Once we repeat the process for $\forall h\in\{1,\ldots,H\}$ and $\forall\alpha\in\{\alpha_1,\ldots,\alpha_A\}$ a pseudo-population of $N=\sum_{h\in\{1,\ldots,H\}}\sum_{\alpha\in\{\alpha_1,\ldots,\alpha_A\}}N_{h,\alpha}=\sum_{j\in S_{\text{ESIE}}} w_j$ units we generated with the information of response variables ($Y$, among others) and strata (hence, information fo the design variables $X_1$, $X_2$ and $X_3$ will also be generated).

Finally we generate the rest of the ovariates as follows $\forall d\in\{4,\ldots,p\}$ assume that $X_d$ is a categorical variable with a total of $K$ categories: $\{1,\ldots,K\}$ categories. Then, for each unit $i$ generated in the pseudo-population ($\forall i \in U$), we generated $x_{di}\in\{1,\ldots,K\}$ following a categorical distribution (i.e., $x_{di}\sim Cat(\pi_{d1},\ldots,\pi_{dK})$) where the probability corresponding to each category $k\in\{1,\ldots,K\}$ is calculated as follows based on the known ESIE finite population $U_{\text{ESIE}}$. Let us assume that $i\in U_{ESIE,h}, \forall h\in\{1,\ldots,H\}$. Then,
\begin{equation}
\pi_{dk}=\dfrac{\sum_{j\in U_{\text{ESIE}}} 1_{U_{\text{ESIE},\:h}}(j)\left[x_{dj}=k\right]}{\sum_{j\in U_{\text{ESIE}}} 1_h(j)}, \forall k\in\{1,\ldots,K\},
\end{equation}
where $1_{U_{\text{ESIE},\:h}}(j)$ is defined in \eqref{indicator_function} and,
\begin{equation}
\left[x_{dj}=k\right]=\left\{
\begin{array}{ccc}
1 & \text{if} & x_{dj}=k,\\
0 & \text{if} & x_{dj}\neq k,
\end{array}
\quad \forall j\in U_{\text{esie}} \text{ and } \forall k\in\{1,\ldots,K\}.
\right.
\end{equation}
In this way, the pseudo-population based on the ESIE survey has been generated with the response variable $Y$, the vector of explanatory variables $\pmb X$ and the strata.

\section{Pseudo-population sampling process}\label{sampling}

The two pseudo-populations have been sampled by one-step stratified sampling, in the same way as the real survey data described in Section \ref{datasets}. 

In order to sample the pseudo-population of the Scenario 1, first, we identify how many units have been sampled from a stratum $h$, $\forall h\in\{1,\ldots,H\}$ in the real survey sample $S_{\text{ESIE}}$ (let us denote this amount as $n_h$). Then, we sample $n_h$ units randomly from stratum $h$ from the pseudo-population $U$. Repeating the same process for $\forall h\in\{1,\ldots,H\}$ we obtain a sample $S.$

Finally, sampling weights are assigned to each sampled unit as follows. For $\forall i\in S$ assume that $i$ is a unit from stratum $h$, $\forall h\in\{1,\ldots,H\}$, then:
\begin{equation}
w_{i}=\dfrac{N_h}{n_h},
\end{equation}
where $N_h$ indicates the number of units in the stratum $h$ in $U$, and $n_h$ the number of units in the stratum $h$ in $S$.

\end{document}